\documentclass[%
 reprint,
superscriptaddress,
 amsmath,amssymb,
 aps,
]{revtex4-1}

\usepackage{graphicx}% Include figure files
\usepackage{dcolumn}% Align table columns on decimal point
\usepackage{bm}% bold math
\usepackage{hyperref}% add hypertext capabilities
\begin{document}

\preprint{APS/123-QED}

\title{Test the weak cosmic censorship conjecture via black hole in dark matter halo
}%
\author{Meirong Tang}
 
\author{Lai Zhao}

\author{Zhaoyi Xu}
 \email{Electronic address: zyxu@gzu.edu.cn(Corresponding author)
}

% \email{Second.Author@institution.edu}
\affiliation{%
College of Physics, Guizhou University,\\
 Guiyang 550025, China
}%

%\collaboration{MUSO Collaboration}%\noaffiliation

%\author{Charlie Author}
 %\homepage{http://www.Second.institution.edu/~Charlie.Author}

%\author{Delta Author}

%\collaboration{CLEO Collaboration}%\noaffiliation

%\date{\today}% It is always \today, today,
             %  but any date may be explicitly specified

\begin{abstract}
The weak cosmic censorship conjecture states that the black hole singularity is hidden inside the event horizon of the black hole, making it impossible for an external observer to measure.
In this study, we investigate the weak cosmic censorship conjecture test of dark matter halo-black hole systems in both the cold dark matter model and ultralight dark matter model scenarios, with the aim of gaining insights into the influence of dark matter particles on the weak cosmic censorship conjecture.
By examining the particle incident on an extremely or nearly extremal dark matter- black hole, as well as the scattering of a scalar field by an extreme or near-extreme dark matter- black hole.
We find that, for test particles, the weak cosmic censorship conjecture is violated under extreme conditions. Under near-extreme conditions, our calculation results show a second-order small quantity, which indicates that the weak cosmic censorship conjecture may be breached under near-extreme conditions. However, if the self-force effect is taken into consideration, whether the weak cosmic censorship conjecture will be violated still requires further in-depth research. For scalar fields, the weak cosmic censorship conjecture is violated under extreme conditions, while under near-extreme conditions, our results show that the weak cosmic censorship conjecture still holds. 
This research will contribute to furthering our comprehension of the intricate interplay between dark matter and black holes.
\begin{description}
\item[Keywords]
Weak cosmic censorship conjecture, Cold dark matter model, Ultralight dark matter model, Black hole, Scalar field
\end{description}
\end{abstract}
%\keywords{Suggested keywords}%Use showkeys class option if keyword

%\arxivnumber{...}
\maketitle

%\tableofcontents
\section{Introduction}
\label{intro}

In 1915, Einstein formulated his elegant theory of general relativity, providing a geometric interpretation of gravity as the curvature of space-time\cite{1972Steven Weinberg-book, 1984Wald-book, 1970Misner-book, 1973Hawking-book}. 
However, the emergence of both Schwarzschild's and Kerr's solutions indicates the fact that spacetime has pathological properties of singularities, characterized by singularitiesian curvature scalar becomes infinitely large. These singularities correspond to an infinite density of matter and cause all observed quantities to diverge\cite{1973Hawking-book}.
Hawking and Penrose employed a global approach to establish the renowned singularity theorem, demonstrating that black hole singularities are an inevitable consequence of gravitational collapse under specific energy conditions\cite{1973Hawking-book, 1965PhRvL..14...57P, 1970RSPSA.314..529H}.
The presence of singularities implies that general relativity is incomplete, thereby providing a strong impetus for its expansion into areas such as string theory and loop quantum gravity\cite{1998Polchinski-book1,1998LRR.....1....1R}. 
The existence of naked singularities, which are not concealed within the event horizon of a black hole, is particularly intriguing among the array of singularities. These naked singularities can be observed by both infinite and finite distance observers. However, when an observer measures the naked singularity, it leads to a divergence in any physical quantity, thus making the physical prediction lose causality.

The cosmic censorship conjecture was proposed by Penrose to mitigate the physical challenges arising from the emergence of singularities\cite{1969NCimR...1..252P, 1979grae.book.....H}.The cosmic censorship conjecture can be classified into two categories: the strong cosmic censorship conjecture and the weak cosmic censorship conjecture.
The strong cosmic censorship conjecture posits the impossibility of time-like singularities and, in more rigorous mathematical terms, can be interpreted as Cauchy horizon (or causal horizon) within black holes that exhibit instability under general space-time perturbations.
The weak cosmic censorship conjecture posits that, subject to specific energy conditions, the space-time resulting from gravitational collapse will remain concealed within the event horizon, rendering it imperceptible to observers outside said horizon. The mathematical intricacy of the cosmic censorship conjecture renders it an ongoing enigma yet to be resolved.

In the following, we focus on the weak cosmic censorship conjecture. Although the weak cosmic censorship conjecture cannot be proved mathematically, physicists have developed indirect methods to test the weak cosmic censorship conjecture.
Currently, there are two approaches employed for testing the weak cosmic censorship conjecture. The first method involves numerical simulations, such as modeling the gravitational collapse of a massive star during its late-stage evolution\cite{1984CMaPh..93..171C, 1991PhRvL..66..994S, 1993PhRvL..70....9C, 1987PhRvL..59.2137O}.
The second approach involves employing an ideal experiment to examine the weak cosmic censorship conjecture. One prominent ideal experiment, proposed by Wald \cite{2019PhRvD.100d4043N, 1974AnPhy..82..548W}, revolves around directing a particle possessing significant angular momentum towards an extreme or near-extreme black hole and subsequently investigating whether the event horizon of said black hole can be disrupted following the particle's impact.
In a series of studies on ideal experiments, physicists have discovered instances where the weak cosmic censorship conjecture is violated\cite{1999PhRvD..59f4013H, 2013PhRvD..87d4028G, 2009PhRvL.103n1101J, 2017PhRvD..96f4010R, 2015GReGr..47..150C}. For instance, Veronika E. Hubeny's calculations demonstrate that by carefully selecting motion parameters such as the mass and angular momentum of the incident particle, it is possible to disrupt the event horizon of a near-extreme Reissner-Nordstrom black hole, leading to the formation of naked singularity \cite{1999PhRvD..59f4013H}.
The weak cosmic censorship conjecture does not appear to be violated in the case of second-order perturbations when injecting high-angular-momentum particles into an extreme Kerr-Newman black hole\cite{2017PhRvD..96j4014S}.

On the other hand, physicists have also tested the weak cosmic censorship conjecture by incident a scalar field on a black hole. The weak cosmic censorship conjecture is not violated when considering the scattering of classical scalar fields on extreme or near-extreme black holes\cite{2021JCAP...10..012G, 2021NuPhB.96515335L, 2020arXiv200402902G, 2020JCAP...03..058G, 2018JHEP...09..081G}. 
However, when considering the superradiation induced by the interaction between the black hole and the scalar field, both extreme and near-extreme black holes situation do not violate the weak cosmic censorship conjecture\cite{2020PhRvL.124u1101B, 2015LNP...906.....B}. Additionally, physicists have discovered that incorporating quantum effects can also lead to a violation of this conjecture, resulting in the emergence of an undesirable naked singularity\cite{2009PhRvD..79j1502M}.
In a real universe, ideal Kerr black holes, Reissner-Nordstrom black holes, and Kerr-Newman black holes cannot exist because black holes cannot exist in a world without an external matter distribution.
Numerous astronomical observations have indicated the existence of an exceptionally intricate distribution of matter in proximity to black holes, and the reality of black holes is contingent upon the consideration of external matter distribution\cite{1973IAUS...55..155S, 1993ARA&A..31..473A}. Of particular interest are cases involving dark matter particles and dark energy surrounding black holes\cite{2005PhR...405..279B, 2018RvMP...90d5002B, 2006IJMPD..15.1753C, 2011CoTPh..56..525L}. The solutions to Einstein's gravitational field equations become exceedingly complex when accounting for the coupling between dark matter and black holes.
In recent years, several approximate solutions have been derived\cite{2018JCAP...09..038X, 2020PhRvD.101b4029X, 2019PhLB..795....1K}. By utilizing these black hole metrics, one can investigate the impact of dark matter on testing the weak cosmic censorship conjecture. Recently, Meng,Xu and Tang discovered that by considering the interaction between an ideal fluid dark matter and a black hole\cite{2023arXiv230812913M}, the weak cosmic censorship conjecture can be violated.

It is intriguing to observe a violation of the weak cosmic censorship conjecture in an perfect fluid dark matter-black hole system, as the presence of dark matter near a black hole necessitates alterations to its spacetime structure and properties. Consequently, understanding the intrinsic characteristics of black holes relies on comprehending violations of the weak cosmic censorship conjecture in dark matter- black hole systems. However, Meng et al. dark matter model that lacks specific dark matter models or dark matter particles, thereby limiting our comprehension of the weak cosmic censorship conjecture. 
In order to incorporate the information of dark matter models and dark matter particles into the test of the weak cosmic censorship conjecture, it is necessary to investigate the test of the weak cosmic censorship conjecture for dark matter-black hole systems under different dark matter models.

In this work, our primary focus is to investigate the impact of cold dark matter and ultralight dark matter on the test of the weak cosmic censorship conjecture for Kerr black holes. The \ref{metric} section presents the spacetime metric for cold dark matter-black hole and ultralight-black hole systems, followed by a calculation of their respective properties.
In Section \ref{test-test particle}, we test the weak cosmic censorship conjecture by injecting particles into the cold dark matter-black hole system and the ultralight dark matter-black hole system. In Section \ref{test-scalar field}, we test the weak cosmic censorship conjecture by injecting classical scalar fields into cold dark matter-black hole systems and ultralight dark matter-black hole systems. The final section is devoted to a summary and discussion. Throughout the paper, we will use the natural system of units $G=c=1$.

\section{The spacetime metric of a rotating black hole in the presence of dark matter}
\label{metric}
The density distribution of dark matter halos in galaxies can be characterized by a few parameters, such as scale radius and critical density, which arise from the interaction among dark matter particles.
If a black hole is located at the geometric center of a dark matter halo, a series of intricate interactions will ensue between the black hole and the surrounding dark matter halo.
The dark matter-black hole systems can be accurately described using the steady-state approximate black hole metric.
The cold dark matter halos determined by the cold dark matter model and the ultralight dark matter halos determined by the ultralight dark matter model are considered in this work.

The density distribution of the dark matter halo in the cold dark matter model is characterized by the Navarro-Frenk-White (NFW) profile \cite{1991ApJ...378..496D, 1997ApJ...490..493N, 1996ApJ...462..563N}, which can be mathematically expressed as follows
\begin{equation}
\rho_{\rm{NFW}}=\frac{\rho_{s}}{\frac{r}{R_{s}}\big(1+\frac{r}{R_{s}}\big)^{2}},
\label{metric1}
\end{equation}
the critical density of the dark matter halo is denoted as $\rho_{s}$, while the scale radius of the dark matter halo is represented by $R_{s}$. In a cold dark matter halo, the spacetime metric of a rotating black hole is \cite{2018JCAP...09..038X}
%\begin{widetext}
\begin{align}
&ds^{2}=-\left[1-\dfrac{r^{2}+2Mr-r^{2}\left(1+\dfrac{r}{R_{s}}\right)^{-\dfrac{8\pi\rho_{s}R_{s}^{3}}{r}}}{\Sigma^{2}}\right]dt^{2}+\notag\\
&\frac{\Sigma^{2}}{\Delta_{CDM}}+\Sigma^{2}d\theta^{2}+\dfrac{\sin^{2}\theta}{\Sigma^{2}}\left[ (r^{2}+a^{2})^{2}-a^{2}\Delta_{CDM}\sin^{2}\theta \right]d\phi^{2}\notag\\
&+\dfrac{2\left[ r^{2}+2Mr-r^{2}\left(1+\dfrac{r}{R_{s}}\right)^{-\dfrac{8\pi\rho_{s}R_{s}^{3}}{r}} \right]a\sin^{2}\theta}{\Sigma^{2}}d\phi dt,
\label{metric2}
\end{align}
%\end{widetext}
the expressions for the functions $\Sigma^{2}$ and $\Delta_{CDM}$ are as follows
\begin{equation}
\Sigma^{2}=r^{2}+a^{2}\cos^{2}\theta,
\label{metric3}
\end{equation}
\begin{equation}
\Delta_{CDM}=r^{2}\left[ 1+\dfrac{r}{R_{s}} \right]^{-\dfrac{8\pi\rho_{s}R_{s}^{3}}{r}}-2Mr+a^{2}.
\label{metric4}
\end{equation}

For the Bose-Einstein condensed dark matter model, the density distribution of the dark matter halo is described by the Thomas-Fermi distribution (TF) \cite{2002CQGra..19.6259U, 2011JCAP...05..022H}. The mathematical form of TF distribution is as follows
\begin{equation}
\rho_{\rm{TF}}=\rho_{c}\frac{\sin(\dfrac{R_{c}r}{\pi})}{\dfrac{R_{c}r}{\pi}},
\label{metric5}
\end{equation}
the center density of the dark matter halo is denoted as $\rho_{c}$, while the scale radius of the dark matter halo is represented by $R_{c}$. In a Bose-Einstein condensed dark matter halo, the spacetime metric of a rotating black hole is \cite{2018JCAP...09..038X}
\begin{align}
&ds^{2}=-\left[ 1-\dfrac{r^{2}+2Mr-r^{2}\exp\left( -\dfrac{8\rho_{c}R^{2}_{c}}{\pi}\dfrac{\sin(\pi r/R_{c})}{\pi r/R_{c}} \right)}{\Sigma^{2}} \right]dt^{2}\notag\\
&+\dfrac{\Sigma^{2}}{\Delta_{ULDM}}dr^{2}+\Sigma^{2}d\theta^{2}+\notag\\
&\dfrac{2\left[ r^{2}+2Mr-r^{2}\exp\left( -\dfrac{8\rho_{c}R^{2}_{c}}{\pi}\dfrac{\sin(\pi r/R_{c})}{\pi r/R_{c}} \right)\right]a\sin^{2}\theta}{\Sigma^{2}}d\phi dt\notag\\
&+\dfrac{\sin^{2}\theta}{\Sigma^{2}}\left[ (r^{2}+a^{2})^{2}-a^{2}\Delta_{ULDM}\sin^{2}\theta \right]d\phi^{2},
\label{metric6}
\end{align}
the expression for the function $\Delta_{ULDM}$ is as follows
\begin{equation}
\Delta_{ULDM}=r^{2}\exp\left( -\dfrac{8\rho_{c}R^{2}_{c}}{\pi}\dfrac{\sin(\pi r/R_{c})}{\pi r/R_{c}} \right)-2Mr+a^{2}.
\label{metric7}
\end{equation}

For the rotating spacetime metric (\ref{metric2}) and (\ref{metric6}), they describe the cold dark matter halo-black hole system and the ultralight dark matter halo-black hole system, respectively. 
When the dark matter halo is absent, meaning that the critical density (or central density) of dark matter halo is zero, the spacetime metric (\ref{metric2}) and (\ref{metric6}) revert to the Kerr black hole scenario.
As long as the critical density (or central density) of dark matter halo is paired with the scale radius (their values are determined by the galactic dark matter halo observation data), the black hole metric in the center of dark matter halo of different galaxies can be obtained.

The main method of testing the weak cosmic censorship conjecture is to find out whether the event horizon of black hole can be destroyed by particle incident or scalar field, so it is necessary to calculate and discuss the event horizon of dark matter-black hole system.
In order to distinguish the different dark matter models, we will use CDM and ULDM to label the two models. From the definition of the event horizon of a rotating black hole, the event horizon of a black hole is given by the following equation
\begin{equation}
g^{\mu\nu}\dfrac{\partial f}{\partial x^{\mu}}\dfrac{\partial f}{\partial x^{\nu}}=0,
\label{metric8}
\end{equation}
where $g^{\mu\nu}$ is the inverse matrix of the black hole spacetime metric, and $f=f(x^{\mu})=0$ is the three-dimensional surface (or hypersurface) in the four-dimensional spacetime.
Equation (\ref{metric8}) shows that the length of the normal vector on this three-dimensional surface is zero, and this equation gives the event radius of the dark matter-black hole spacetime.
In the case of axisymmetry, where the function f is a function of both $r$ and $\theta$, equation (\ref{metric8}) can be expressed in the following manner
\begin{equation}
g^{11}\left(\dfrac{\partial f}{\partial r}\right)^{2}+g^{22}\left(\dfrac{\partial f}{\partial \theta}\right)^{2}=0.
\label{metric9}
\end{equation}
Through simple calculation, the result is as follows
\begin{equation}
g^{rr}=0  \quad  or \quad  \dfrac{\Delta_{r}}{\Sigma^{2}}=0.
\label{metric10}
\end{equation}
The specific equation for the event horizon structure in the cold dark matter model is as follows
\begin{equation}
r^{2}\left[1+\dfrac{r}{R_{s}}\right]^{-\dfrac{8\pi\rho_{s}R_{s}^{3}}{r}}-2Mr+a^{2}=0.
\label{metric11}
\end{equation}
The specific equation for the event horizon structure in the ultralight dark matter model is as follows
\begin{equation}
r^{2}\exp\left[ -\dfrac{8\rho_{c}R^{2}_{c}}{\pi}\dfrac{\sin(\pi r/R_{c})}{\pi r/R_{c}} \right]-2Mr+a^{2}=0.
\label{metric12}
\end{equation}

\subsection{Cold dark matter model scenario}
\label{cdm-pro}
It can be seen from equation (\ref{metric11}) that there is no exact analytical solution for this equation. However, considering that the influence of dark matter on the space - time structure is usually relatively weak, approximate solutions can be gradually obtained through iterative methods. For this purpose, we start from the analytical solution of the event horizon of the classical Kerr black hole. In the standard Kerr metric, the position of the event horizon is given by the following analytical expression:
\begin{equation}
r_{h}^0=M\pm\sqrt{M^2-a^2}.
\label{ker1}
\end{equation}
After taking into account the influence of the cold dark matter halo, the position of the event horizon will be modified based on this standard solution. By using the iterative method, the first-order approximate solution can be obtained as
\begin{widetext}
\begin{equation}
r_{h}^{\mathrm{I}}=M\pm \sqrt{M^2-a^2+\left(M\pm \sqrt{M^2-a^2}\right)^2 \left(1-\left(\frac{M\pm \sqrt{M^2-a^2}}{R_s}+1\right)^{-\frac{8
  \pi\rho _s R_s^3  }{M\pm \sqrt{M^2-a^2}}}\right)}   .
\label{cold-1}
\end{equation}
\end{widetext}
On this basis, the second-order iterative solution \(r_{h}^{\mathrm{II}}\) can be further written as
\begin{equation}
r_{h}^{\mathrm{II}}=M\pm \sqrt{M^2-a^2+{r_{h}^{\mathrm{I}}}^{2} \left(1-\left(\frac{r_{h}^{\mathrm{I}}}{R_s}+1\right)^{-\frac{8 \pi \rho_s R_s^3 
   }{r_{h}^{\mathrm{I}}}}\right)}.
\label{cold-2}
\end{equation}
Theoretically, higher-order approximate solutions can be obtained through further iterations. However, in the discussion of this paper, we only consider the second-order iterative approximation, and the error of this approximation is analyzed in Appendix \ref{A} to evaluate its rationality.

To simplify the subsequent derivation process, we introduce a parameter \(k_{1}\), which is defined as 
\begin{equation}
k_{1}=8 \pi \rho _s R_s^3 ,
\label{cold-3}
\end{equation}
Therefore, the expression for the event horizon can be rewritten as
\begin{equation}
r_{h}^{\mathrm{II}}=M\pm \sqrt{M^2-a^2+{r_{h}^{\mathrm{I}}}^{2} \left(1-\left(\frac{r_{h}^{\mathrm{I}}}{R_s}+1\right)^{-\frac{k_{1}
   }{r_{h}^{\mathrm{I}}}}\right)}.
\label{cold-4}
\end{equation}

The $+$ sign in analogy to the Kerr black hole represents the event horizon of the dark matter-black hole, while the $-$ sign corresponds to the Cauchy event horizon (also known as causal event horizon) of the dark matter-black hole.
The changing behaviour and structure of the dark matter-black hole event horizon depend on the model parameters of the NFW profile, specifically, the critical density $\rho_{s}$ and scale radius $R_{s}$, as evidenced by combining equations (\ref{cold-3}) and (\ref{cold-4}).
When $\rho_{s}=0$ or $R_{s}\rightarrow\infty$, i.e. there are no cold dark matter particles near the black hole, the black hole event horizon (\ref{cold-4}) retreats to the Kerr black hole case.
When condition $a^{2}\leq M^{2}$ is satisfied, the spacetime metric (\ref{metric2}) describes a spinning black hole at the center of dark matter halo.
When the condition $a^{2}>M^{2}$ is satisfied, the spacetime metric (\ref{metric2}) describes a rotating spacetime with no event horizon.

The surface area of the event horizon of a cold dark matter-black hole is
\begin{equation}
A=\int\int\sqrt{g_{\theta\theta}g_{\phi\phi}}d\theta d\phi.
\label{metric16}
\end{equation}
Combined with the spacetime metric (\ref{metric2}), there is
\begin{equation}
g_{\theta\theta}=\Sigma^{2},
\label{metric17}
\end{equation}
\begin{equation}
g_{\phi\phi}=\dfrac{\sin^{2}\theta}{\Sigma^{2}}\left[ (r_{H}^{2}+a^{2})^{2}-a^{2}\Delta_{CDM}\sin^{2}\theta \right],
\label{metric18}
\end{equation}
\begin{equation}
g_{\theta\theta}g_{\phi\phi}=\Sigma^{2}\cdot\dfrac{\sin^{2}\theta}{\Sigma^{2}}\left[ (r_{h}^{2}+a^{2})^{2}-a^{2}\Delta_{CDM}\sin^{2}\theta \right]  $$$$
=\sin^{2}\theta\left[ (r_{H}^{2}+a^{2})^{2}-a^{2}\times 0\times\sin^{2}\theta \right]$$$$
=\sin^{2}\theta\left(r_{h}^{2}+a^{2}\right)^{2},
\label{metric19}
\end{equation}
the surface area of the event horizon of the black hole is as follows
\begin{align}
A=\int\int\sqrt{g_{\theta\theta}g_{\phi\phi}}d\theta d\phi 
&=\int\int\sin\theta(r_{h}^{2}+a^{2})d\theta d\phi \notag \\
&=4\pi(r_{h}^{2}+a^{2}).
\label{metric20}
\end{align}
In order to calculate the Hawking temperature of a dark matter-black hole, it is necessary to first calculate the apparent gravity of the black hole
\begin{align}
k_{+}&=\lim_{r \to r_{h}}\left[ -\dfrac{1}{2}\sqrt{\dfrac{g^{rr}}{-\hat{g}_{tt}}}\hat{g}_{tt,r} \right]\notag\\
&=\lim_{r \to r_{h}}\left[ -\dfrac{1}{2}\sqrt{\dfrac{g^{rr}}{\left( \dfrac{g_{t\phi}^{2}}{g_{\phi\phi}}-g_{tt} \right)}}\cdot\left( g_{tt}-\dfrac{g_{t\phi}^{2}}{g_{\phi\phi}} \right)_{,r} \right] \notag\\
&=\dfrac{r_{h}-r_{-}}{2(r_{h}^{2}+a^{2})}=\dfrac{1}{2}\dfrac{r_{h}^{2}-a^{2}}{r_{h}(r_{h}^{2}+a^{2})}.
\label{metric21}
\end{align}
In black hole thermodynamics, the steady-state black hole has a Hawking temperature, which is proportional to the apparent gravity of the black hole, so the Hawking temperature can be obtained
\begin{equation}
T_{h}=\dfrac{k_{+}}{2\pi}=\dfrac{r_{h}^{2}-a^{2}}{4\pi r_{h}(r_{h}^{2}+a^{2})}.
\label{metric22}
\end{equation}
The rotational velocity of the black hole can be determined by calculating the metric coefficient of the dark matter-black hole spacetime, and if the radius of the event horizon is considered, the angular velocity is
\begin{align}
\Omega_{h}=&-\dfrac{g_{t\phi}}{g_{\phi\phi}}=-\dfrac{\left[ r_{h}^{2}+2Mr_{h}-r^{2}\left( 1+\dfrac{r_{h}}{R_{s}} \right)^{-\dfrac{8\pi\rho_{s}R_{s}^{3}}{r_{h}}} \right]a\sin^{2}\theta}{\Sigma^{2}}\notag\\
&\times \dfrac{\Sigma^{2}}{\sin^{2}\theta\left[ (r_{h}^{2}+a^{2})^{2}-a^{2}\Delta_{CDM}\sin^{2}\theta \right]}\notag\\
&=\dfrac{a}{r_{h}^{2}+a^{2}}.
\label{metric23}
\end{align}

\subsection{Ultralight dark matter model scenario}
\label{uldm-pro}
For black holes with an ultralight dark matter halo, the expression for their event horizons can refer to the analysis method of the cold dark matter model (see Subsection \ref{cdm-pro}). Starting from equation (\ref{metric12}), since the dark matter density \(\rho_{c} \ll 1\), the first-order analytical approximation can be written as
\begin{widetext}
\begin{equation}
r_{h}^{\mathrm{I}}=M\pm \sqrt{M^2-a^2+\left(M\pm \sqrt{M^2-a^2}\right)^2 \left(1-\exp \left(-\frac{8 \rho _c R_c^3 \sin \left(\frac{\pi
    \left(M\pm \sqrt{M^2-a^2}\right)}{R_c}\right)}{\pi ^2 \left(M\pm
   \sqrt{M^2-a^2}\right)}\right)\right)} .
\label{ult1}
\end{equation}
\end{widetext}
The corresponding second-order analytical approximation is
\begin{equation}
r_{h}^{\mathrm{II}}=M\pm \sqrt{M^2-a^2+{r_{h}^{\mathrm{I}}}^2 \left(1-e^{-\frac{8 \rho _c R_c^3 \sin \left(\frac{\pi 
   r_{h}^{\mathrm{I}}}{R_c}\right)}{\pi ^2 r_{h}^{\mathrm{I}}}}\right)}.
\label{ult2}
\end{equation}
By iterating in sequence, higher-order analytical solutions can be obtained. The analysis in this paper mainly focuses on the second-order analytical solutions, and the corresponding error analysis is presented in  Appendix \ref{A}.

To simplify the subsequent derivation process, we introduce a parameter \(k_{2}\), which is defined as
\begin{equation}
k_{2}=\frac{8 \rho _c R_c^3}{\pi ^2}.
\label{ult3}
\end{equation}
Therefore, the expression for the event horizon can be rewritten as
\begin{equation}
r_{h}^{\mathrm{II}}=M\pm \sqrt{M^2-a^2+{r_{h}^{\mathrm{I}}}^2 \left(1-e^{-\frac{k_2 \sin \left(\frac{\pi 
   r_{h}^{\mathrm{I}}}{R_c}\right)}{r_{h}^{\mathrm{I}}}}\right)}.
\label{ult4}
\end{equation}

The analysis of the event horizon structure of the ultralight dark matter halo-Kerr black hole bears resemblance to that of the cold dark matter halo-black hole (similar to the analysis after equation (\ref{cold-4})), and the corresponding results can also be obtained.
The main findings are as follows. The event horizon $r_{h}$ of the ultralight dark matter halo-Kerr black hole exhibits a strong dependence on the model parameters of the ULDM halo, specifically, the central density $\rho_{c}$ and the parameter $R_{c}$.
When the central density $\rho_{c}=0$, i.e., the absence of ULDM halo, leads to the degeneration of $r_{h}$ into the Kerr black hole scenario.
When condition $a^{2}\leq M^{2}$ is satisfied, the spacetime metric (\ref{metric6}) characterizes a rotating black hole that encodes information regarding the interaction between ULDM and Kerr black holes. When condition $a^{2}>M^{2}$ is fulfilled, the spacetime metric (\ref{metric6}) describes a rotating spacetime devoid of an event horizon.

The surface area $A$, surface gravity $k_{+}$, Hawking temperature $T_{h}$, and angular velocity of the event horizon $\Omega_{h}$ of the ultra-light dark matter halo-black hole can be determined using calculations similar to those presented in the previous section. Specifically, it is only necessary to substitute the event horizon radius of the ultra-light dark matter halo-black hole for $r_{h}$ in formulas (\ref{metric20}), (\ref{metric21}), (\ref{metric22}) and (\ref{metric23}).

\section{Test the weak cosmic censorship conjecture by incident particles into a cold dark matter-black hole and ultralight dark matter-black hole}
\label{test-test particle}
We will now examine the weak cosmic censorship conjecture of cold dark matter halo-Kerr black holes and ultralight dark matter halo-Kerr black holes.
The discussion will commence by examining the event horizon structure of a black hole. By applying equations (\ref{cold-4}) and (\ref{ult4}), we can derive the subsequent deformation
\begin{equation}
\frac{r_{h}^{\mathrm{II}}}{M}=\left\{ \begin{aligned}
& 1\pm \sqrt{\frac{M^2-a^2+{r_{h}^{\mathrm{I}}}^{2} \left(1-\left(\frac{r_{h}^{\mathrm{I}}}{R_s}+1\right)^{-\frac{k_{1}
   }{r_{h}^{\mathrm{I}}}}\right)}{M^2}}\; CDM \\
&1\pm \sqrt{\frac{M^2-a^2+{r_{h}^{\mathrm{I}}}^2 \left(1-e^{-\frac{k_2 \sin \left(\frac{\pi 
   r_{h}^{\mathrm{I}}}{R_c}\right)}{r_{h}^{\mathrm{I}}}}\right)}{M^2}}\; ULDM \\
\end{aligned}\right.
\label{part27}
\end{equation}
Since the expression of \(r_{h}^{\mathrm{I}}\) in the above formula contains \(\sqrt{M^{2}-a^{2}}\), therefore, when the condition \(a^{2}\leq M^{2}\) is satisfied, both the space-time metrics (\ref{metric2}) and (\ref{metric6}) possess at least one black hole event horizon.
%when condition $a^{2}+k_{i}\leq M^{2}$ is satisfied, both the space-time metrics (\ref{metric2}) and (\ref{metric6}) possess at least one black hole event horizon.
When the condition $a^{2}> M^{2}$ is satisfied, the space-time metrics (\ref{metric2}) and (\ref{metric6}) lack a black hole event horizon, thereby exposing the singularity within the black hole, transforming it into a naked singularity.
According to the physical discussion presented in (\ref{part27}), the condition for destroying the event horizon of the black hole is as follows
\begin{equation}
a^{2}>M^{2},
\label{part28}
\end{equation}
The black hole spin represents the angular momentum per unit mass of the black hole, which is denoted as $a=J/M$. Therefore, inequality (\ref{part28}) can be rewritten as follows

\begin{equation}
J>M^{2}.
\label{part30}
\end{equation}

To further elucidate the space-time properties and their underlying physics within the event horizons of both cold and ultra-light dark matter halo-Kerr black holes, a test particle with significant angular momentum can be injected into these types of black holes. 
When the test particle is incident, it forms a new system with the central rotating black hole. As a result, the original rotating black hole alters the event horizon structure of the new system by interacting with the test particle. This interaction may involve changes in the mass and angular momentum of the new black hole system, ultimately determining whether the event horizon of the new black hole can be destabilized. If the event horizon of the new black hole can be destroyed, the angular momentum and mass of the new system must satisfy specific conditions
\begin{equation}
J^{'}>M^{'2}.
\label{part31}
\end{equation}

In a dark matter-Kerr black hole system, when a test particle of mass m is injected into the black hole, it is known from general relativity that the test particle is moving along the geodesic space-time of the dark matter-Kerr black hole. When the affine parameter on the geodesic is selected as the proper time $\tau$, the geodesic equation of the test particle motion is
\begin{equation}
\dfrac{d^{2}x^{\mu}}{d\tau^{2}}+\Gamma_{\alpha\beta}^{\mu}\dfrac{dx^{\alpha}}{d\tau}\dfrac{dx^{\beta}}{d\tau}=0.
\label{part32}
\end{equation}
The geodesic equation can in fact be derived by varying the Lagrangian quantity of the particle. Theoretical calculations indicate that the Lagrangian quantity of the test particle can be obtained as 
\begin{equation}
L=\dfrac{1}{2}mg_{\mu\nu}\dfrac{dx^{\mu}}{d\tau}\dfrac{dx^{\nu}}{d\tau}=\dfrac{1}{2}mg_{\mu\nu}x^{\mu}x^{\nu}.
\label{part33}
\end{equation}
To facilitate the ensuing calculations, we restrict the test particle to the equatorial plane of the black hole spacetime, with the test particle incident from infinity. 
According to the fundamental characteristics of a rotating black hole, the angular coordinate $\theta$ remains constant over time, and the momentum component in the $\theta$ direction is null, i.e., $P_{\theta}=mg_{22}\dot{\theta}=0$. Therefore, the energy and angular momentum of the test particle are 
\begin{equation}
\delta E=-p_{t}=-\dfrac{\partial L}{\partial t}=-mg_{0\nu}x^{\nu},
\label{part34}
\end{equation}
\begin{equation}
\delta J=p_{\phi}=\dfrac{\partial L}{\partial \phi}=mg_{3\nu}x^{\nu}.
\label{part35}
\end{equation}
When the original dark matter-Kerr black hole accrues test particles, the energy and angular momentum of the resulting black hole are thereby determined. They are
\begin{equation}
M^{'}=M+\delta E,
\label{part36}
\end{equation}
\begin{equation}
J{'}=J+\delta J.
\label{part37}
\end{equation}
We need to find out if the event horizon of a black hole is destroyed when the test particle enters it.
The purpose of this test is to determine whether the particles entering a newly formed black hole satisfy the conditions necessary for surpassing the event horizon. In case these conditions are met, an singularity ring within the black hole becomes visible to an observer located at infinity.

For the test particle with mass m, its four-dimensional velocity satisfies the normalization condition, and its mathematical expression is $U^{\mu}U_{\mu}=g^{\mu\nu}p_{\mu}p_{\nu}/m^{2}=-1$. 
The normalization condition can be simplified by substituting expressions (\ref{part34}) and (\ref{part35}) into this equation. They are
\begin{equation}
\left\{\begin{aligned}
& g^{\mu\nu}p_{\mu}p_{\nu}=-m^{2} \\
& g^{00}\delta E^{2}+g^{11}p_{r}^{2}+g^{22}p_{\theta}^{2}+g^{33}\delta J^{2}-2g^{03}\delta E \delta J=-m^{2}. \\
\end{aligned}\right.
\label{part38}
\end{equation}
The equation set is solved, and the energy change of the test particle is
\begin{equation}
\delta E=\dfrac{g^{03}}{g^{00}}\delta J \pm\dfrac{1}{g^{00}}\sqrt{(g^{03})^{2}\delta J^{2}-g^{00}(g^{33}\delta J^{2}+g^{11}p_{r}^{2}+m^{2})}.
\label{part39}
\end{equation}
Outside the black hole event horizon, geodesics are time-like geodesics that point to the future, so there is $dt/d\tau>0$. Expand the expression (\ref{part34}) and expression (\ref{part35}) into the following form
\begin{equation}
\left\{\begin{aligned}
& mg_{00}\dfrac{dt}{d\tau}+mg_{03}\dfrac{d\phi}{d\tau}=-\delta E \\
& mg_{30}\dfrac{dt}{d\tau}+mg_{33}\dfrac{d\phi}{d\tau}=\delta J. \\
\end{aligned}\right.
\label{part40}
\end{equation}
By solving the binary first order equations, we can get
\begin{equation}
\dfrac{dt}{d\tau}=-\dfrac{g_{03}\delta J+g_{33}\delta E}{g_{00}g_{33}-g_{03}^{2}}.
\label{part41}
\end{equation}
Combined conditions $dt/d\tau>0$ and equation (\ref{part41}), it can be obtained that the energy increment and angular momentum increment of the test particle satisfy the following inequality
\begin{equation}
\delta E>-\dfrac{g_{03}}{g_{33}}\delta J.
\label{part42}
\end{equation}
The negative sign should be considered in equation (\ref{part39}), based on this observation, and its expression is as follows
\begin{equation}
\delta E=\dfrac{g^{03}}{g^{00}}\delta J-\dfrac{1}{g^{00}}\sqrt{(g^{03})^{2}\delta J^{2}-g^{00}(g^{33}\delta J^{2}+g^{11}p_{r}^{2}+m^{2})}.
\label{part43}
\end{equation}
According to equation (\ref{part42}), a lower bound exists for the energy increment of the test particle, as well as an upper limit for its angular momentum increment. The upper limit for the angular momentum increment of the test particle is
\begin{equation}
\delta J<-\lim_{r \to r_{h}^{\mathrm{II}}}\dfrac{g_{33}}{g_{03}}\delta E=\dfrac{{r_{h}^{\mathrm{II}}}^{2}+a^{2}}{a}\delta E=\dfrac{1}{\Omega_{h}}\delta E=\delta J_{max}.
\label{part44}
\end{equation}
Here, we use the second-order analytical solution \(r_{h}^{\mathrm{II}}\) of the event horizon to replace the event horizon \(r_{h}\).
By examining the destroy condition of the event horizon in a rotating black hole $J^{'}>M^{'2}$,it becomes evident that there is a minimal value for the angular momentum of a test particle. Defining $J^{'}=J+\delta J$ as the angular momentum of the new system and $M^{'}=M+\delta E$ as the mass of the new system. If the formula $J^{'}>M^{'2}$ is expanded, the lower limit that the angular momentum increment satisfies is
\begin{equation}
\delta J>(M^{2}-J)+2M\delta E+\delta E^{2}=\delta J_{min}.
\label{part45}
\end{equation}
By combining the expression (\ref{part44}) and (\ref{part45}), the physical conditions for incident particles to destroy the dark matter-black hole event horizon can be obtained
\begin{equation}
\delta J_{min}<\delta J<\dfrac{1}{\Omega_{h}}\delta E.
\label{part46}
\end{equation}

The destruction of the black hole event horizon occurs when the angular momentum increment of the test particle falls within this specified range. Consequently, the ring singularity within the black hole becomes exposed, thereby enabling the violation of the weak cosmic censorship conjecture at this particular moment.

Next, we explore the possibility of the destruction of the event horizon in the extreme dark matter-black hole scenario. The condition satisfied by the extreme black hole is $a^{2}=M^{2}$.In this case, the angular velocity of the black hole is
\begin{widetext}
\begin{align}
\frac{1}{\Omega_{h}}=\dfrac{{r_{h}^{\mathrm{II}}}^{2}+a^{2}}{a}=\left\{ \begin{aligned}
& 2 M+2 \sqrt{k_1}\sqrt{M \ln \left(\frac{M+R_s}{R_s}\right)}+k_1 \left(\frac{M}{M+R_s}+2 \ln \left(\frac{M+R_s}{R_s}\right)\right)+O\left((k_1)^2\right)\quad CDM \\
&2 M+2 \sqrt{k_2} \sqrt{M \sin \left(\frac{\pi  M}{R_c}\right)}+k_2 \left(2 \sin \left(\frac{\pi 
   M}{R_c}\right)+\frac{\pi  M \cos \left(\frac{\pi  M}{R_c}\right)}{R_c}\right)+O\left((k_2)^2\right)\quad ULDM \\
\end{aligned}\right..
\label{part48}
\end{align}
\end{widetext}
So, when analyzing the scenario of dark matter-black hole, the conditions for us to destroy the event horizon of the extreme dark matter-black hole are as follows
\begin{widetext}
\begin{align}
\delta J_{max}-\delta J_{min}=\left\{ \begin{aligned}
&2 \sqrt{k_1}\sqrt{M \ln \left(\frac{M+R_s}{R_s}\right)}\delta E
+k_1 \left(\frac{M}{M+R_s}+2 \ln \left(\frac{M+R_s}{R_s}\right)\right)\delta E-\delta E^2+O\left(k_1^2\right)\\
&2\sqrt{k_2} \sqrt{M \sin \left(\frac{\pi  M}{R_c}\right)}\delta E+k_2 \left(2 \sin \left(\frac{\pi 
   M}{R_c}\right)+\frac{\pi  M \cos \left(\frac{\pi  M}{R_c}\right)}{R_c}\right)\delta E-\delta E^2+O\left(k_2^2\right) \\
\end{aligned}\right..
\label{part49}
\end{align}
\end{widetext}
Apparently, it can be seen from the above formula (\ref{part49}) that since the values of \(k_{1}=8 \pi \rho _s R_s^3 \) and \(k_{2}=\frac{8 \rho _c R_c^3}{\pi ^2}\) are both non-negative (being zero in the absence of dark matter), and \(\sqrt{k_i}\delta E > k_i\delta E \sim  \delta E^2\) (where \(i = 1, 2\)), this formula thus indicates that \(\delta J_{max}-\delta J_{min}\geq 0\), and the equality holds in the absence of dark matter. It can be inferred from this that the existence of dark matter may disrupt the structure of the event horizon of black holes (whether it is an extreme cold dark matter halo-black hole or an extreme ultralight dark matter halo-black hole), while in the absence of dark matter, this structure can remain intact.

Next, we examine the test of the weak cosmic censorship hypothesis for near-extreme black holes, taking the energy increment $\delta E$ of the test particle to first-order approximation. Then, we can deduce that the condition for the destruction of the new black hole event horizon following the particle's entry into the black hole event horizon is as follows
\begin{equation}
\delta J_{min}=2M\delta E+(M^{2}-J)
<\delta J<\delta J_{max}=\dfrac{r_{H}^{2}+a^{2}}{a}\delta E.
\label{part52}
\end{equation}  

For the near-extreme black hole spacetime, a parameter $\varepsilon$ can be used to describe the degree of approximation to the extreme black hole, $\varepsilon$ satisfies the following relation
\begin{equation}
\dfrac{a^{2}}{M^{2}}=1-\varepsilon^{2}.
\label{part53}
\end{equation}
According to the relation (\ref{part53}), when $\varepsilon\rightarrow 0$, the black hole is the near extreme case, and when $\varepsilon=0$, the black hole is the extreme case. Using the relation (\ref{part53}), the condition (\ref{part52}) for destroying the event horizon of the black hole is simplified as

\begin{equation}
\dfrac{1}{\Omega_{h}}-2M-M^{2}(1-\sqrt{1-\varepsilon^{2}})>0.
\label{part54}
\end{equation}

For the case of near-extreme black holes, that is, when the dimensionless parameter \(\varepsilon \to 0\), for the cold dark matter halo-black hole, its result can be obtained by calculating the conditional expression (\ref{part54}) as
\begin{align}
\dfrac{1}{\Omega_{h}}&-2M-M^{2}(1-\sqrt{1-\varepsilon^{2})}=2M\varepsilon\notag\\
&+k_1\left[\frac{M}{M+R_s}+2\ln\left(\frac{M}{M+R_s}\right)+\frac{\ln\left(\frac{M}{M+R_s}\right)}{\varepsilon}\right]\notag\\
&\frac{1}{2}k_1\varepsilon\left[\frac{M(3M+4R_s)}{(M+R_s)^2}+3\ln\left(\frac{M+R_s}{R_s}\right) \right]\notag\\
&+O\left(k_1^2,\varepsilon^2\right);
\label{part55a}
\end{align}
while for the ultralight dark matter halo-black hole, its result can be obtained by calculating the conditional expression (\ref{part54}) as
\begin{align}
&\dfrac{1}{\Omega_{h}}-2M-M^{2}(1-\sqrt{1-\varepsilon^{2}})\notag\\
&=2 M \varepsilon+k_2\left[\frac{\sin \left(\frac{\pi 
   M}{R_c}\right)}{\varepsilon}+2 \sin \left(\frac{\pi  M}{R_c}\right)+\frac{\pi  M
   \cos \left(\frac{\pi  M}{R_c}\right)}{R_c}
 \right]\notag\\
 &+k_2\varepsilon \left[-\frac{\pi ^2 M^2 \sin \left(\frac{\pi 
   M}{R_c}\right)}{2 R_c^2}+\frac{3}{2} \sin \left(\frac{\pi 
   M}{R_c}\right)+\frac{2 \pi  M \cos \left(\frac{\pi 
   M}{R_c}\right)}{R_c}\right]\notag\\
  &+O\left(k_2^2,\varepsilon^2\right).
\label{part55b}
\end{align}

Regarding the results of the above equations (\ref{part55a}) and (\ref{part55b}), when only the low-order approximations are taken into account, the result obtained is a second-order small quantity (here the approximation of \(O\left(\varepsilon\right) \sim  O\left(\delta E \right)\sim  O(k_1\text{or} k_2)\sim O\left(\frac{M}{R_s}\text{or}\frac{M}{R_c} \right)\) is adopted). Such a result forces us to consider the second-order approximation. When considering the second-order approximation, we get
\begin{widetext}
\begin{equation}
\delta J_{max}-\delta J_{min}=\left\{ \begin{aligned}
&M \varepsilon\delta E+ k_1 \delta E\left[\frac{M}{M+R_s}+2\ln\left(\frac{M}{M+R_s}\right)+\frac{\ln\left(\frac{M}{M+R_s}\right)}{\varepsilon}\right]+O(\delta E^3)\quad CDM \\
&M \varepsilon\delta E+k_2 \delta E\left[\frac{\sin \left(\frac{\pi 
   M}{R_c}\right)}{\varepsilon}+2 \sin \left(\frac{\pi  M}{R_c}\right)+\frac{\pi  M
   \cos \left(\frac{\pi  M}{R_c}\right)}{R_c}
 \right]+O(\delta E^3)\quad ULDM \\
\end{aligned}\right..
\label{parta}
\end{equation}
\end{widetext}
%Regarding the result of the above equation, we find that when considering only low-order approximations, the obtained result is a second-order small quantity. This forces us to consider second-order approximations. When considering the second-order approximation, we obtain

Through the analysis of this result, it can be found that obviously, \(\delta J_{\text{max}} - \delta J_{\text{min}} > 0\), and thus the condition (\(\ref{part54}\)) is satisfied. From a mathematical perspective, it seems to suggest that both cold dark matter and ultralight dark matter may lead to the destruction of the event horizon of black holes. However, further analysis shows that the lowest-order contribution of this result is a second-order quantity. When fully considering the space-time background inside black holes and the back-reaction effects of particles, it is found that the possibility of disrupting near-extreme black holes is extremely low. This is due to the fact that the influence generated by the self-force effect of particles or other related effects belongs to the second-order quantity, so these contributions cannot be ignored in the analysis \cite{2017PhRvD..96j4014S}. Future research will further explore the influence of these  effects in depth. Therefore, under these circumstances, whether the cosmic censorship conjecture will be violated remains an open scientific question worthy of in-depth exploration.

\section{Test the weak cosmic censorship conjecture by incident scalar field into a cold dark matter-black hole and ultralight dark matter-black hole}
\label{test-scalar field}
Semiz et al. (2011) proposed that a scalar field with large angular momentum can be used to scatter extreme or near-extreme black holes in order to destroy the black hole event horizon \cite{2011GReGr..43..833S}. In this section, we scatter a scalar field over the extreme or near extreme dark matter-black hole event horizon, calculate the probability of event horizon destruction in the extreme or near extreme dark matter-black hole case, and try to understand how dark matter changes the weak cosmic censorship conjecture test problem.

Then we calculate the scalar field scattering black hole and obtain the energy increment and angular momentum increment of the system after the scalar field incident.

There exists a scalar field $\psi$ in the dark matter-black hole space-time with a corresponding particle mass $\mu$. Then the interaction between the scalar field and the black hole is described by the Klein-Gordon equation, whose mathematical form is
\begin{equation}
\nabla_{\nu}\nabla^{\nu}\psi-\mu^{2}\psi=0,
\label{scalar56}
\end{equation}
where $\nabla_{\nu}$ is the Laplacian operator under the dark matter-black hole space-time metric. By substituting it into the Klein-Gordon equation, the Klein-Gordon equation can be written in a more concrete form
\begin{equation}
\dfrac{1}{\sqrt{-g}}\partial_{\mu}(\sqrt{-g}g^{\mu\nu}\partial_{\nu}\psi)-\mu^{2}\psi=0,
\label{scalar57}
\end{equation}
where $g$ is the determinant of the dark matter-black hole space-time metric, and $g^{\mu\nu}$ is the inverse matrix of the dark matter-black hole space-time metric. According to the space-time line elements (\ref{metric2}) and (\ref{metric6}), the corresponding inverse metric matrix is obtained by calculation
\begin{equation}
g^{00}=-\dfrac{(r^{2}+a^{2})^{2}-a^{2}\Delta_{i}\sin^{2}\theta}{\Delta_{i}\Sigma^{2}},  \quad\quad\quad\quad   g^{11}= \dfrac{\Delta_{i}}{\Sigma^{2}},  $$$$
g^{33}=\dfrac{\Delta_{i}-a^{2}\sin^{2}\theta}{\Delta_{i}\Sigma^{2}\sin^{2}\theta},  \quad\quad\quad\quad   g^{22}=\dfrac{1}{\Sigma^{2}},  $$$$
g^{03}=g^{30}=\dfrac{-a}{\Delta_{1}\Sigma^{2}}\times $$$$
\dfrac{r^{2}+2Mr-r^{2}\left( 1+\dfrac{r}{R_{s}} \right)^{-\dfrac{8\pi\rho_{s}R_{s}^{3}}{r}}}{k_{1}} \quad CDM $$$$
g^{03}=g^{30}=\dfrac{-a}{\Delta_{2}\Sigma^{2}}\times $$$$
 \dfrac{r^{2}+2Mr-r^{2}\exp\left( -\dfrac{8\rho_{c}R^{2}_{c}}{\pi}\dfrac{\sin(\pi r/R_{c})}{\pi r/R_{c}} \right)}{k_{2}}. \quad ULDM,
\label{scalar58}
\end{equation}
where $i=1,2$,  for the cold dark matter model, $\Delta_{1}=\Delta_{CDM}$; for the ultralight dark matter model, $\Delta_{2}=\Delta_{ULDM}$.
These inverse matrix components and the metric determinant $g$ are substituted into the Klein-Gordon equation (\ref{scalar57}), thereby reducing the Klein-Gordon equation to a more concrete form. For cold dark matter-black holes, the corresponding equation is
\begin{equation}
-\dfrac{(r^{2}+a^{2})^{2}-a^{2}\Delta_{1}\sin^{2}\theta}{\Delta_{1}\Sigma^{2}}\dfrac{\partial^{2}\psi}{\partial t^{2}}-\dfrac{2ak_{1}}{\Delta_{1}\Sigma^{2}}\dfrac{\partial^{2}\psi}{\partial t \partial\phi}+$$$$
\dfrac{1}{\Sigma^{2}}\dfrac{\partial}{\partial r}\left( \Delta_{1}\dfrac{\partial\psi}{\partial r} \right) 
+\dfrac{1}{\Sigma^{2}\sin\theta}\dfrac{\partial}{\partial \theta}\left( \sin\theta\dfrac{\partial\psi}{\partial \theta} \right) 
$$$$
+\dfrac{\Delta_{1}-a^{2}\sin^{2}\theta}{\Delta_{1}\Sigma^{2}\sin\theta}\dfrac{\partial^{2}\psi}{\partial\phi^{2}}-\mu^{2}\psi=0.
\label{scalar59}
\end{equation}
For ultralight dark matter- black holes, the corresponding equation is
\begin{equation}
-\dfrac{(r^{2}+a^{2})^{2}-a^{2}\Delta_{2}\sin^{2}\theta}{\Delta_{2}\Sigma^{2}}\dfrac{\partial^{2}\psi}{\partial t^{2}}-\dfrac{2ak_{2}}{\Delta_{2}\Sigma^{2}}\dfrac{\partial^{2}\psi}{\partial t \partial\phi}+$$$$
\dfrac{1}{\Sigma^{2}}\dfrac{\partial}{\partial r}\left( \Delta_{2}\dfrac{\partial\psi}{\partial r} \right)
+\dfrac{1}{\Sigma^{2}\sin\theta}\dfrac{\partial}{\partial \theta}\left( \sin\theta\dfrac{\partial\psi}{\partial \theta} \right) $$$$
+\dfrac{\Delta_{2}-a^{2}\sin^{2}\theta}{\Delta_{2}\Sigma^{2}\sin\theta}\dfrac{\partial^{2}\psi}{\partial\phi^{2}}-\mu^{2}\psi=0.
\label{scalar60}
\end{equation}
The equations (\ref{scalar59}) and (\ref{scalar60}) are rather intricate and cannot be solved directly. Nonetheless, they can be analyzed employing the classical method of separation of variables, resulting in the wave function 
$\psi$ being expressed in the following product form
\begin{equation}
\psi(t,r,\theta,\phi)=e^{-i\omega t}R(r)S_{lm}(\theta)e^{im\phi},
\label{scalar61}
\end{equation}
where $\omega$ represents the scalar field oscillation frequency, $R(r)$ is the radial unknown function, $S_{lm}(\theta)$ is the spherical harmonic function, and $l$ and $m$ are the angular quantum number and magnetic quantum number respectively. The equation (\ref{scalar61}) is substituted for the entrance equation (\ref{scalar59}) and the field equation (\ref{scalar60}) to separate the variables, and the following form is obtained
\begin{align}
&\dfrac{(r^{2}+a^{2})^{2}-a^{2}\Delta_{i}\sin^{2}\theta}{\Delta_{i}\Sigma^{2}}\omega^{2}e^{-i\omega t}R(r)S_{lm}(\theta)e^{im\phi}\notag\\
&-\dfrac{2ak_{i}}{\Delta_{i}\Sigma^{2}}m\omega e^{-i\omega t}R(r)S_{lm}(\theta)e^{im\phi}+\notag\\
&\dfrac{1}{\Sigma^{2}}e^{-i\omega t}S_{lm}(\theta)e^{im\phi}\dfrac{d}{dr}\left(\Delta_{i}\dfrac{dR(r)}{dr}\right)+\notag\\
&\dfrac{1}{\Sigma^{2}\sin\theta}e^{-i\omega t}R(r)e^{im\phi}\dfrac{d}{d\theta}\left(\sin\theta\dfrac{dS_{lm}(\theta)}{d\theta}\right) \notag\\
&+\dfrac{a^{2}\sin^{2}\theta-\Delta_{i}}{\Delta_{i}\Sigma^{2}\sin\theta}e^{-i\omega t}R(r)S_{lm}(\theta)m^{2}e^{im\phi}-\notag\\
&\mu^{2}e^{-i\omega t}R(r)S_{lm}(\theta)e^{im\phi}=0.
\label{scalar62}
\end{align}  
In the expression (\ref{scalar62}), the term with angle $\theta$ can be separated from the term with radial distance $r$, thus decomposing the equation (\ref{scalar62}) into two equations. The spherical harmonic 
$S_{lm}(\theta)$ satisfies the following angular equation
\begin{equation}
\dfrac{1}{\sin\theta}\dfrac{d}{d\theta}\left[ \sin\theta\dfrac{dS_{lm}(\theta)}{d\theta} \right]-$$$$
\left[ a^{2}\omega^{2}\sin^{2}\theta+\dfrac{m^{2}}{\sin^{2}\theta}+\mu^{2}a^{2}\cos^{2}\theta-\lambda_{lm} \right]S_{lm}=0,
\label{scalar63}
\end{equation}
the radial function $R(r)$ satisfies the following radial equation
\begin{align}
&\dfrac{d}{dr}\left( \Delta_{i}\dfrac{dR}{dr} \right)+\notag\\
&\left[ \dfrac{(r^{2}+a^{2})^{2}}{\Delta_{i}}\omega^{2}-\dfrac{2ak_{i}}{\Delta_{i}}m\omega+\dfrac{m^{2}a^{2}}{\Delta_{i}}-\mu^{2}r^{2}-\lambda_{lm} \right]R(r)=0.
\label{scalar64}
\end{align}
The solution of equation (\ref{scalar63}) represents a spherical harmonic function, where the constant $\lambda_{lm}$ serves as the eigenvalue for the spherical harmonic function. Owing to the normalization property of the spherical function, the integral of this function evaluates to unity when calculating the energy flux over the event horizon. Consequently, it does not contribute significantly to our estimation of the energy flux.
The solution of the angular equation is accordingly factored into our computation, followed by addressing the radial equation, which constitutes the more delicate aspect of this computation. Given the singularity of coordinates at the black hole event horizon in the BL coordinate system, it becomes imperative to adopt a more suitable coordinate system to proceed with our calculation.
The Eddington-Fingelstein coordinate system, commonly referred to as the turtle coordinate, is typically employed in the standard calculation process. Generally, the turtle coordinate and the BL coordinate are related as follows
\begin{equation}
\dfrac{dr}{dr_{*}}=\dfrac{\Delta_{i}}{r^{2}+a^{2}}.
\label{scalar65}
\end{equation}
In this coordinate system, turtle coordinates effectively cover all regions of space-time both inside and outside the dark matter - black hole. Substituting this spatiotemporal coordinate transformation into the radial equation (\ref{scalar64}) transforms the equation (\ref{scalar64}) into the following form
\begin{equation}
\dfrac{\Delta_{i}}{(r^{2}+a^{2})^{2}}\dfrac{d}{dr}(r^{2}+a^{2})\dfrac{dR}{dr_{*}}+\dfrac{d^{2}R}{dr_{*}^{2}}+\left( \omega-\dfrac{ma}{r^{2}+a^{2}} \right)^{2}R(r)$$$$
+\left[\dfrac{\Delta_{i}}{(r^{2}+a^{2})^{2}}2am\omega-\dfrac{\Delta_{i}}{(r^{2}+a^{2})^{2}}(\mu^{2}r^{2}+\lambda_{lm})\right]R(r)=0.
\label{scalar66}
\end{equation}
According to the fundamental framework of dark matter-black hole space-time scattering via scalar field, it becomes imperative to compute the energy and angular momentum fluxes at the event horizon subsequent to the scalar field's entry into the black hole's event horizon. This necessitates the adoption of the radial radius in proximity to the event horizon. Consequently, the condition $\Delta_{i}\approx 0$ must be satisfied. In this scenario, the radial equation (\ref{scalar66}) can be simplified into the following form
\begin{equation}
\dfrac{d^{2}R}{dr_{*}^{2}}+\left( \omega-\dfrac{ma}{r_{h}^{2}+a^{2}} \right)^{2}R=0.
\label{scalar67}
\end{equation}
The rewritten equation (\ref{scalar67}) based on the angular velocity (\ref{metric23}) at the dark matter-black hole event horizon can be expressed as follows
\begin{equation}
\dfrac{d^{2}R}{dr_{*}^{2}}+( \omega- m\Omega_{h})^{2}R=0.
\label{scalar68}
\end{equation}
The general solution of this second-order ordinary differential equation is as follows
\begin{equation}
R(r)=\exp[\pm i(\omega-m\Omega_{h})r_{*}].
\label{scalar69}
\end{equation}
The plus sign corresponds to the outgoing wave after the scattering of the scalar field and the black hole, while the minus sign corresponds to the incident wave after the scattering. According to the physical requirements of this study, the incident wave should be naturally selected. Combining the separate variable expression of the scalar field wave function, the approximate form of the scalar field near the dark-black hole event horizon can be derived
\begin{equation}
\psi(t,r,\theta,\phi)=\exp[- i(\omega-m\Omega_{h})r_{*}]e^{-i\omega t}S_{lm}(\theta)e^{im\phi}.
\label{scalar70}
\end{equation}
The expression (\ref{scalar70}) elucidates the specific configuration acquired by a scalar field in proximity to the event horizon when it interacts with a dark matter-black hole. The structure of this expression is dictated by the interplay between the scalar field and the black hole. By employing expression (\ref{scalar70}), we can compute the energy flux and angular momentum flux subsequent to the scalar field's incidence on the black hole event horizon. With this information, we can assess whether the event horizon of the dark-black hole spacetime can be sustained subsequent to the scalar field's incidence.

The expression (\ref{scalar70}) indicates that the quantum numbers $l$ and $m$ describe the characteristics of the scalar field as it crosses the event horizon of a black hole. Consequently, $(l,m)$ can be utilized to identify the mode of the scalar field. When scattering a scalar field $(l,m)$ with dark matter-black holes, the interaction between the dark matter-black hole and the scalar field results in the black hole event horizon both reflecting and absorbing portions of the scalar field. 
The portion of the absorbed scalar field can be transmuted into the energy and angular momentum of the dark matter- black hole.

To compute the alteration of energy and angular momentum for the dark matter black hole when the scalar field impinges on it, it's imperative to determine the energy- momentum tensor via the scalar field wave function $\psi$. The expression for the energy-momentum tensor is as follows
\begin{equation}
T_{\mu\nu}=\partial_{\mu}\psi \partial_{\nu}\psi^{*}-\dfrac{1}{2}g_{\mu\nu}(\partial_{\mu}\psi \partial^{\nu}\psi^{*}+\mu^{2}\psi^{*}\psi).
\label{scalar71}
\end{equation}  
In this energy-momentum tensor, the component $T_{\quad t}^{r}$ determines the energy flux of the scalar field, and the component $T_{\quad \phi}^{r}$ determines the angular momentum flux of the scalar field. Through calculation, it is found that the energy flux of the scalar field near the dark matter-black hole event horizon is
\begin{equation}
\dfrac{dE}{dt}=\int\limits_{h} T_{\quad t}^{r}\sqrt{-g}d\theta d\phi=\omega(\omega-m\Omega_{h})(r_{h}^{2}+a^{2}),
\label{scalar72}
\end{equation}
the angular momentum flux of the scalar field near the dark matter-black hole event horizon is
\begin{equation}
\dfrac{dJ}{dt}=\int\limits_{h} T_{\quad \phi}^{r}\sqrt{-g}d\theta d\phi=m(\omega-m\Omega_{h})(r_{h}^{2}+a^{2}).
\label{scalar73}
\end{equation}
For the dark matter-black hole case determined by different dark matter models, it is only necessary to replace the event horizon radius and angular velocity in the expressions (\ref{scalar72}) and (\ref{scalar73}). When condition $\omega>m\Omega_{H}$ is satisfied, the rotating dark-matter black hole gains energy and angular momentum from the scalar field. According to (\ref{scalar72}) and (\ref{scalar73}), the energy and angular momentum acquired by the black hole in a short time are
\begin{equation}
dE=\omega(\omega-m\Omega_{H})(r_{H}^{2}+a^{2})dt
\label{scalar74}
\end{equation}
and
\begin{equation}
dJ=m(\omega-m\Omega_{H})(r_{H}^{2}+a^{2})dt.
\label{scalar75}
\end{equation}
The expressions (\ref{scalar74}) and (\ref{scalar75}) can be used to calculate the incremental energy and angular momentum of the rotating dark matter-black hole. The problem of whether the black hole event horizon can be destroyed when the scalar field scatters the black hole can be analyzed.

We project a monochromatic scalar field (with frequency $\omega$ and quantum number $m$) onto a rotating dark matter- black hole, aiming to determine whether this scalar field can disrupt the event horizon of the black hole. This concept bears resemblance to that of particles incident upon a black hole causing destruction of its event horizon.
The initial mass of the rotating dark matter-black hole system is $M$ and the initial angular momentum is $J$ before the monochromatic scalar field incident. When the monochromatic scalar field is incident, the mass of the rotating dark matter-black hole becomes $M'=M+dE$ and the angular momentum becomes $J'=J+dJ$.
According to the dark matter- black hole event horizon destruction condition, whether the event horizon of the complex system is destroyed after the monochromatic scalar field incident on the black hole depends on the sign of $M^{'2}-J^{'}$. When $M^{'2}-J^{'}<0$ is satisfied, the event horizon of the rotating dark matter-black hole is destroyed by a monochromatic scalar field. Otherwise it will not be destroyed.

When a very short time interval $dt$ is taken, the monochromatic scalar field will cause the mass of the rotating dark matter-black hole to increase $dE$ and angular momentum to increase $dJ$. The $M^{'2}-J^{'}$ of the rotating dark matter-black hole system after absorbing a monochromatic scalar field is

\begin{equation}
M^{'2}-J^{'}=(M^{2}-J)+2MdE+d E^2-dJ.
\label{scalar76}
\end{equation}

Ignoring the higher order term of $dE$, by substituting (\ref{scalar74}) and (\ref{scalar75}) into (\ref{scalar76}), (\ref{scalar76}) becomes the following form
\begin{align}
M^{'2}-J^{'}=&(M^{2}-J)+2M m^{2}\left( \dfrac{\omega}{m}-\dfrac{1}{2M}\right)
\left( \dfrac{\omega}{m}-\Omega_{h} \right)\notag\\
&\times(r_{h}^{2}+a^{2})dt.
\label{scalar77}
\end{align}

When the rotating dark matter - black hole becomes an extreme black hole, that is, when the condition $J=M^{2}$ is satisfied, $M^{'2}-J^{'}$ will become a simpler form as follows

\begin{align}
M^{'2}-J^{'}=2M m^{2}\left( \dfrac{\omega}{m}-\dfrac{1}{2M}\right)
\left( \dfrac{\omega}{m}-\Omega_{H} \right)(r_{h}^{2}+a^{2})dt.
\label{scalar78}
\end{align}

At the same time, when the spinning dark matter-black hole reaches the extreme, the angular velocity of the event horizon becomes
%\begin{widetext}
\begin{align}
&\Omega_{h}=\dfrac{a}{{r_{h}^{\mathrm{II}}}^{2}+a^{2}}\notag\\
&=\left\{ \begin{aligned}
&\frac{1}{2 M} -\frac{\sqrt{k_1} \sqrt{M \ln \left(\frac{M}{R_c}+1\right)}}{2 M^2}-\frac{k_1}{4 M
   \left(R_c+M\right)}\quad CDM \\
&\frac{1}{2 M}-\frac{\sqrt{k_2} \sqrt{M \sin \left(\frac{\pi  M}{R_c}\right)}}{2 M^2}-\frac{\pi  k_2 \cos
   \left(\frac{\pi  M}{R_c}\right)}{4 M R_c}\quad ULDM \\
\end{aligned}\right..
\label{scalar79}
\end{align}
%\end{widetext}
According to the expression (\ref{scalar78}) for the extreme black hole case $M^{'2}-J^{'}$, if the incident scalar field satisfies the following conditions
\begin{equation}
\dfrac{\omega}{m}=\frac{1}{2}\left( \dfrac{1}{2M}+\Omega_{h} \right),
\label{scalar80}
\end{equation}  
then $M^{'2}-J^{'}$ further becomes the following form

\begin{equation}
M^{'2}-J^{'}=-\dfrac{1}{2}M m^{2}
\left( \Omega_{h}-\dfrac{1}{2M} \right)^{2}(r_{h}^{2}+a^{2})dt.
\label{scalar81}
\end{equation}
Apparently, by combining expressions (\ref{scalar79}) and (\ref{scalar81}), it can be concluded that both the extreme cold dark matter halo-black hole and the extreme ultralight dark matter halo-black hole can satisfy \(M^{'2}-J^{'}\leq 0\). This means that the existence of dark matter may cause the event horizon of these black holes to be disrupted by the scalar field. When there is no dark matter, this inequality degenerates into an equality, and the spacetime metric at this time corresponds to the situation of an extreme Kerr black hole. It is worth noting that this conclusion is consistent with the theoretical result that the event horizon of an extreme Kerr black hole cannot be destroyed by the scalar field.

The destruction of event horizons in the case of near-extreme black holes has been previously discussed (particle incidence) and is also a topic of interest. It poses an intriguing question whether a scalar field can disrupt the event horizon when incident upon a near-extreme rotating dark matter-black hole spacetime. In the near-extreme scenario, $J$ does not equal $M^{2}$, thereby altering the circumstances.

As discussed earlier, event horizon destruction in the case of near-extreme black holes is also of interest. Whether the event horizon can be disrupted by a scalar field is a very interesting question when considering a scalar field incident into near-extreme rotating dark matter-black hole spacetime. In the near-extreme black hole case, $J$ is not equal to $M^{2}$, at this point $M^{'2}-J^{'}$ becomes the following
\begin{align}
M^{'2}-J^{'}=&(M^{2}-J)+2M m^{2}\left( \dfrac{\omega}{m}-\dfrac{1}{2M}\right)
\left( \dfrac{\omega}{m}-\Omega_{h} \right)\notag\\
&\times(r_{h}^{2}+a^{2})dt.
\label{scalar83}
\end{align}

If the incident scalar field mode is restricted as follows
\begin{equation}
\dfrac{\omega}{m}=\dfrac{1}{2}\left( \Omega_{h}+\dfrac{1}{2M} \right),
\label{scalar84}
\end{equation}
the expression (\ref{scalar83}) becomes the following
\begin{align}
M^{'2}-J^{'}=&(M^{2}-J)-\dfrac{1}{8M}m^{2}
\Omega_{h}^{2}\left( \dfrac{1}{\Omega_{h}}-2M\right)^{2}\notag\\
&\times(r_{h}^{2}+a^{2})dt.
\label{scalar85}
\end{align}

Here, we adopt the method of analyzing test particles in Section \ref{test-test particle} to define a dimensionless parameter \(\varepsilon\) that describes the degree of deviation from the extreme case. Its expression is \(\dfrac{a^2}{M^2}=1-\varepsilon^2\). Since \(\varepsilon\) approaches \(0\), the Taylor expansion of $M^{2}-J$ is carried out, and the result is 
\begin{equation}
M^{2}-J=\frac{M^2\varepsilon^{2}}{2}-O(\varepsilon^{3}),
\label{scalar86}
\end{equation}
the Taylor expansion is also performed on $\dfrac{1}{\Omega_{h}}-2M$, and the result is
\begin{widetext}
\begin{align}
&\dfrac{1}{\Omega_{h}}-2M
=\left\{\begin{aligned}
&2M\varepsilon
+k_1\left[\frac{M}{M+R_s}+2\ln\left(\frac{M}{M+R_s}\right)+\frac{\ln\left(\frac{M}{M+R_s}\right)}{\varepsilon}\right]\\&\quad\quad\quad+\frac{1}{2}k_1\varepsilon\left[\frac{M(3M+4R_s)}{(M+R_s)^2}+3\ln\left(\frac{M+R_s}{R_s}\right) \right]+O(\varepsilon^2,k_1^2)\quad\quad\quad\quad\quad CDM \\
&2 M \varepsilon+k_2\left[\frac{\sin \left(\frac{\pi 
   M}{R_c}\right)}{\varepsilon}+2 \sin \left(\frac{\pi  M}{R_c}\right)+\frac{\pi  M
   \cos \left(\frac{\pi  M}{R_c}\right)}{R_c}
 \right]\\
 &\quad\quad\quad+k_2\varepsilon \left[-\frac{\pi ^2 M^2 \sin \left(\frac{\pi 
   M}{R_c}\right)}{2 R_c^2}+\frac{3}{2} \sin \left(\frac{\pi 
   M}{R_c}\right)+\frac{2 \pi  M \cos \left(\frac{\pi 
   M}{R_c}\right)}{R_c}\right]+O(\varepsilon^2,k_2^2)\quad ULDM \\
\end{aligned}\right..
\label{scalar87}
\end{align}
\end{widetext}

Combining formulas (\ref{scalar85}), (\ref{scalar86}) and (\ref{scalar87}), since the time interval \(dt\), the parameter \(\varepsilon\), and the dark matter parameters \(k_1\) or \(k_2\) are all first-order small quantities, it can be obtained that: in the absence of dark matter, the spacetime degenerates into a near-extreme Kerr black hole, and the result at this time is \(M^{'2}-J^{'}>0\). This indicates that the event horizon of the near-extreme Kerr black hole cannot be destroyed by the scalar field, and this result is consistent with the conclusion that the event horizon of the near-extreme Kerr black hole cannot be destroyed by the scalar field. In the presence of dark matter, regardless of the influence of the cold dark matter halo or the light dark matter halo, the result is still \(M^{'2}-J^{'}>0\). This further demonstrates that when considering the coupling effect between the near-extreme Kerr black hole and dark matter, the event horizon of the near-extreme Kerr black hole remains stable and will not be destroyed by the scalar field. Therefore, this analysis result further supports the weak cosmic censorship conjecture.

\begin{table*}[]
    \centering
    \renewcommand{\arraystretch}{1.5} % 调整行间距，数值越大行间距越大
    \caption{For the cold dark matter halo-black hole system, the table lists the analytical approximations from the zeroth order to the third order, the exact numerical solutions and their corresponding errors. The parameters are fixed as \( M = 1\), \( \rho_s =0. 1\), \(R_s= 5\). The calculation expression for the error is
\(\text{Error} = \frac{\lvert r^{\text{numerical}} - r^{\text{iterative}} \rvert}{r^{\text{numerical}}}\).}
    \begin{tabular}{p{1.0cm}p{1.7cm}p{1.7cm}p{1.8cm}p{1.7cm}p{1.8cm}p{1.7cm}p{1.8cm}p{1.7cm}p{1.7cm}}
        \hline\hline % 创建表格顶部的双线
  {Spin}     & \multicolumn{1}{l}{Num. Sol.} & \multicolumn{2}{l}{0th-Ord. Corr.} & \multicolumn{2}{l}{1st-Ord. Corr.} & \multicolumn{2}{l}{2nd-Ord. Corr.} & \multicolumn{2}{l}{3rd-Ord. Corr.}
 \\
        \hline
        $a$ &  $r_{+}^{num}$ &  $r_{+}^0$ & Error &  $r_{+}^{\mathrm{I}}$ & Error & $r_{+}^{\mathrm{II}}$ & Error &  $r_{+}^{\mathrm{III}}$ & Error \\
        \hline
0&2.033845& 2.000000&0.016641&2.032827&0.000500 &2.033814&0.000015&	2.033844 &0.000000 
 \\
0.5&1.900391&1.866025 &0.018083&1.899267&0.000591 &1.900354 &0.000019&1.900389&	0.000001 
 \\
0.9&1.477404&1.435890 &0.028099&1.475324 &0.001408&1.477298 &0.000071 &1.477398 &0.000004 
\\
0.99&1.214664 &1.141067 &0.060590&1.207716&	0.005720 &1.214002&0.000545 &1.214601 &0.000052 
\\
0.999&1.161364&	1.044710&0.100446 &1.147116 &0.012269&	1.159625 &	0.001498 &1.161152& 	0.000183 
 \\
0.9999&1.154868 &1.014142 &0.121855 &1.136964&	0.015503&	1.152600 &0.001964 &1.154581 &	0.000249 \\
\hline\hline % 创建表格底部的双线
    \end{tabular}
    \label{tab1}
\end{table*}
\begin{table*}[]
    \centering
    \renewcommand{\arraystretch}{1.5} % 调整行间距，数值越大行间距越大
    \caption{
  For the  ultralight  dark matter halo - black hole system, the table lists the analytical approximations from the zeroth order to the third order, the exact numerical solutions and their corresponding errors. The parameters are fixed as \( M = 1\), \( \rho_c = 0.1\), \(R_c= 5\). The calculation expression for the error is
\(\text{Error} = \frac{\lvert r^{\text{numerical}} - r^{\text{iterative}} \rvert}{r^{\text{numerical}}}\).}
    \begin{tabular}{p{1.0cm}p{1.7cm}p{1.7cm}p{1.8cm}p{1.7cm}p{1.8cm}p{1.7cm}p{1.8cm}p{1.7cm}p{1.7cm}}
        \hline\hline % 创建表格顶部的双线
  {Spin} & \multicolumn{1}{l}{Num. Sol.} & \multicolumn{2}{l}{0th-Ord. Corr.} & \multicolumn{2}{l}{1st-Ord. Corr.} & \multicolumn{2}{l}{2nd-Ord. Corr.} & \multicolumn{2}{l}{3rd-Ord. Corr.}
\\
       \hline
        $a$ &  $r_{+}^{num}$ &  $r_{+}^0$ & Error &  $r_{+}^{\mathrm{I}}$ & Error & $r_{+}^{\mathrm{II}}$ & Error &  $r_{+}^{\mathrm{III}}$ & Error \\
\hline
0&2.094571&	2.000000&0.045151&2.088926 &	0.002695 &2.094240&	0.000158 &2.094552 	&0.000009 \\
0.5&1.964940 &1.866025&	0.050340&	1.958026 &0.003518&1.964462 &0.000243&	1.964907 	&0.000017 
\\
0.9&1.562616 &1.435890&0.081099&1.547403 &0.009736&	1.560792&	0.001167 &1.562397 &	0.000140 
\\
0.99&1.342747 &1.141067&0.150199 &1.304566&	0.028435 &1.335608&0.005317 &1.341416 &	0.000991 
 \\
0.999&1.308612&1.044710&0.201666 &1.252842&0.042618 &1.297126&	0.008777&1.306261& 	0.001797 
\\
0.9999&1.304922&1.014142 &0.222833 &1.242489&0.047844 &1.291910&0.009972&1.302229 	&0.002064 \\
\hline\hline % 创建表格底部的双线
    \end{tabular}
    \label{tab2}
\end{table*}

\section{Summary}
\label{discuss}
The weak cosmic censorship conjecture is one of the fundamental hypotheses concerning the nature of black holes. Its primary concept posits that there exist certain universal principles aimed at concealing black hole singularities (or ring singularity), thereby preserving space-time causality to a significant degree.
The weak cosmic censorship conjecture imposes a stringent constraint on the solution of the gravitational field equation within the framework of general relativity.
The weak cosmic censorship conjecture currently lacks mathematical proof, but physicists have employed indirect methodologies, such as particle and scalar field incidence, to experimentally examine its validity.
If the weak cosmic censorship conjecture is not violated, then the causality of space-time is preserved. If the weak cosmic censorship conjecture is violated, then the physical difficulties posed by space-time singularity will prompt physicists to revise general relativity.

The present study employs particle incidence and scalar field incidence to examine the weak cosmic censorship conjecture in the context of rotating dark matter-black hole, aiming to investigate whether this conjecture is violated when considering the interaction between Kerr black holes and dark matter halos.
The focus of our analysis lies in examining the potential disruption of the event horizon of a rotating dark matter-black hole under extreme and near-extreme circumstances. The following is our discovery. 
(1) When we incident the test particle on a rotating dark matter-black hole, in the extreme dark matter-black hole case, the dark matter-black hole event horizon can be destroyed, and the weak cosmic censorship conjecture is violated. 
In the near-extreme dark matter-black hole case. Our calculations have revealed the presence of a second-order small quantity capable of disrupting the event horizon of a near-extreme black hole. However, as stated in literature \cite{2017PhRvD..96j4014S} , at the second-order level, it is necessary to fully account for the spacetime background, self-force effects, and other effects, which means considering all the second-order energy contributions related to particle parameters. This implies that further exploration is needed to determine whether the event horizon of a near-extreme black hole can be disrupted when these effects are fully considered.
(2) When we incident a scalar field on a dark matter-black hole, in the case of an extreme dark matter-black hole, the event horizon can be destroyed by the scalar field. In the case of a near-extreme dark matter-black hole, the event horizon will not be destroyed, and the weak cosmic censorship conjecture is not violated.

The results of our calculations align with those of Meng et al.\cite{2023arXiv230812913M}, with the exception that their work focuses on ideal fluid dark matter-black holes, whereas our study examines cold dark matter-black holes and ultralight dark matter-black hole systems. Therefore, the study here will go further.
Our results show that the weak cosmic censorship conjecture is the same for both the cold dark matter particle-black hole interaction system ($k_{1}=8 \pi \rho _s R_s^3$ case) and the ultralight dark matter particle-black hole system ($k_{2}=\frac{8 \rho _c R_c^3}{\pi ^2}$ case).That is, in these two systems, the existence of dark matter causes the event horizon to be destroyed in certain cases.

The present study examines the weak cosmic censorship conjecture of cold dark matter halo-black hole systems and ultralight dark matter halo-black hole systems. However, the model lacks sufficient details regarding the interaction between dark matter particles and black holes, necessitating further investigation in future research.

\begin{acknowledgments}
We acknowledge the anonymous referee for a constructive report that has significantly improved this paper. This work was supported by the Special Natural Science Fund of Guizhou University (Grant No.X2022133), the National Natural Science Foundation of China (Grant No. 12365008) and the Guizhou Provincial Basic Research Program (Natural Science) (Grant No. QianKeHeJiChu-ZK[2024]YiBan027).
\end{acknowledgments}

\section{Appendix}
\label{A}
\subsection{Error evaluation}
In our analysis, since the coupling between dark matter and the black hole is relatively weak, that is \(\rho_s\ll 1\) and \(\rho_c\ll 1\). To obtain an approximation of the analytical expression for the radius of the black hole's event horizon, we use an iterative solution method to derive the corresponding analytical solution. Specifically, we start from the standard Kerr black hole and gradually iterate on its basis to successively obtain the event horizon radius that includes the dark matter effect. In this paper, we use the simplified analytical solution iterated to the second order for analysis. Theoretically, through multiple iterations, a more accurate value of the event horizon radius can be obtained. However, to simplify the analysis and maintain the controllability of the calculation, this paper only uses the analytical solution of the second-order iteration. To evaluate the rationality of this method, we verified the results through numerical solution and calculated the corresponding error to ensure the accuracy and reliability of the results.

In Table \ref{tab1} and Table \ref{tab2}, we respectively present the analytical approximation solutions and iterative solutions of the first three orders under specific parameters for the cold dark matter halo-black hole and the ultralight dark matter halo-black hole, and also give the corresponding errors. Apparently, the error of the second-order analytical approximation solution adopted in this paper is within the range of one thousandth. Such precision fully demonstrates that it is reasonable and reliable to adopt the second-order analytical approximation in the analysis of this paper.

%\bibliographystyle{apsrev4-1}

%\clearpage % 强制分页，确保表格在参考文献之前

\end{document}